\documentclass[12pt,preprint]{aastex}

\begin{document}

%%%%%%%%%%%%%% MY DEFINITIONS
\def\pdot {\dot P}
\def\Omdot {\dot \Omega}
\def\ltsima{$\; \buildrel < \over \sim \;$}
\def\lsim{\lower.5ex\hbox{\ltsima}}
\def\gtsima{$\; \buildrel > \over \sim \;$}
\def\gsim{\lower.5ex\hbox{\gtsima}}
\def\msole{~M_{\odot}}
\def\mdot {\dot M}
\def\sgr {SGR~1806--20}
\def\cha {\textit{Chandra~}}
\def\xmm  {\textit{XMM-Newton~}}
%%%%%%%%%%%%%%%%%%%%%%%%%%%%%%%%%%%%%%

 \slugcomment{Submitted to ApJLett on 28/2/2005 - revised 1/4
}

\title{The first giant flare from \sgr:\\
 observations with the INTEGRAL
SPI Anti-Coincidence Shield}
\author{
S. Mereghetti$^1$,
D. G\"{o}tz\footnote{Istituto di Astrofisica Spaziale e Fisica Cosmica, Sezione di Milano  ''G.Occhialini'' - INAF v.Bassini 15, I-20133 Milano, Italy},
A. von Kienlin$^2$, 
A. Rau$^2$, 
G. Lichti\footnote{Max-Planck-Institut f\"{u}r extraterrestrische Physik, Giessenbachstrasse,  Postfach 1312, D-85741 Garching, Germany},
G. Weidenspointner$^3$, 
P. Jean\footnote{Centre d'\'{E}tude Spatiale des Rayonnements, 31028 Toulouse, France }
}

\begin{abstract}

A giant flare from the Soft Gamma-ray Repeater \sgr\ has been
discovered with the INTEGRAL gamma-ray observatory on 2004
December 27 and detected by many other satellites. This tremendous
outburst, the first one observed from this source, was a hundred
times more powerful than the two giant flares previously observed
from other Soft Gamma-ray Repeaters (SGR). The 50 ms resolution
light curve obtained with the Anticoincidence Shield of the
INTEGRAL  spectrometer SPI, which provides a high effective area
above $\gsim$80 keV, shows evidence for emission lasting about one
hour after the start of the outburst. This component, which decays
in time as $\sim$t$^{-0.85}$, could be the first detection of a
hard X-ray afterglow  associated to an SGR giant flare. The short
(0.2 s) initial pulse was so strong to saturate the detector for
$\sim$0.7 s, and its backscattered radiation from the Moon was
detected 2.8 s later. The following $\sim$400 s long tail,
modulated at the neutron star rotation period of 7.56 s, had a
fluence of 2.6$\times10^{-4}$ erg cm$^{-2}$ above 80 keV, which
extrapolating to lower energies corresponds to an emitted energy
of 1.6$\times10^{44}$~d$^{2}_{15 kpc}$ erg at E$>$3 keV. This is
of the same order of that in the pulsating tails of the two giant
flares seen from other SGRs, despite the hundredfold
larger overall emitted energy of the 2004 December 27 event.
\end{abstract}


\section{Introduction}

Soft Gamma-ray Repeaters (SGRs) are high-energy sources
characterized by sporadic periods of activity in which they emit
short bursts ($<$1~s) with energy up to $\sim$10$^{41}$ erg. They
are believed to be highly magnetized (B$\sim10^{14}$--10$^{15}$ G)
neutron stars, or ''magnetars'' (see Woods \& Thompson 2004 for  a
recent review). Large soft $\gamma$-ray flares, reaching peak
luminosities above several 10$^{44}$ erg s$^{-1}$ and lasting a
few minutes, have been observed from two SGRs: on 1979 March 5
from SGR 0525--66 (Mazets et al. 1979) and on 1998 August 27 from
SGR 1900$+$14 (Hurley et al. 1999; Feroci et al. 1999). The
detection of only two of such giant flares from different sources
in about 30 years implies that these events, involving energy
releases of more than 10$^{44}$ ergs, are relatively rare.

\sgr\ is currently the most prolific of the four known SGRs. Its
level of activity has been increasing in the last few years,
during which many bursts have been detected with different
satellites, including RXTE (G{\" o}{\u g}{\" u}{\c s} et al. 2001;
Ibrahim et al. 2003; Woods et al. 2004) and INTEGRAL (G\"{o}tz et
al. 2004). On 2004 October 5 two clusters of strong bursts with a
total fluence of $\sim$10$^{-4}$ erg cm$^{-2}$  were emitted
within a time span of a few minutes (Mereghetti et al. 2004). It
was noted (Golenetskii et al. 2004) that a similar event occurred
in SGR 1900$+$14 three months before its 1998 giant flare (Aptekar
et al. 2001). The increasing level of activity in \sgr\ was also
reflected in the properties of its quiescent X--ray emission.
XMM-Newton observations showed that its 2-10 keV luminosity
doubled and the spectrum became harder  in 2004 (Mereghetti et al.
2005b). A similar trend was present in the persistent 20-150 keV
emission discovered with INTEGRAL observations carried out in
2003-2004 (Mereghetti et al. 2005a).

The energetic activity of \sgr\ culminated on 2004 December 27,
when the INTEGRAL satellite discovered the first giant outburst
from this source (Borkowski et al. 2004). More than twenty
satellites recorded this exceptional event which started with a
hard pulse so intense to saturate most detectors (Hurley et al.
2005; Mazets et al. 2005; Palmer et al. 2005;  Terasawa et al.
2005) and to significantly ionize the Earth's upper atmosphere
(Campbell et al. 2005). Here we report the properties of this
giant outburst derived with the Anti-Coincidence Shield  (ACS) of
the Spectrometer on board  INTEGRAL (SPI, Vedrenne et al. 2003).

\section{The Anti-Coincidence Shield of SPI}

The ACS consists of 91 bismuth
germanate (BGO) scintillator crystals of thickness between 16 and
50 mm and with a total mass of 512 kg. Besides serving  as a veto
for the SPI germanium spectrometer, the ACS is able to detect
gamma-ray bursts from a large fraction of the sky (von Kienlin et
al. 2003). The data used here consist of the overall ACS count
rate (i.e. the read out from the sum of the 91 crystals) sampled
in time intervals of 50 ms. No energy or directional information
is available. The ACS is sensitive to photons above a low-energy
threshold corresponding to approximately 80 keV. However, due to
the different properties of the various BGO blocks, their
associated photomultipliers, and the chosen redundancy concept
(the signals from two different crystals are fed to the same
front-end electronics), this threshold is not sharp and its exact
value not well determined.

Dead time and saturation effects in the crystals and electronics
are negligible ($<$1\%) for total count rates smaller than a few
10$^5$ cts s$^{-1}$. When this condition is verified we can easily
convert the observed count rates to an incident energy flux,
assuming a spectral shape and knowing the ACS effective area. The
latter is a function of the incidence angle, with a maximum for
directions nearly orthogonal to the satellite pointing axis and
unobstructed by other instruments. At the time of the flare \sgr\
was at a zenith angle $\theta$=106$^{\circ}$ from the SPI pointing
axis and at an azimuth angle $\phi$=9$^{\circ}$
($\phi$=0$^{\circ}$ corresponds to the satellite Sun-pointing
side, i.e. the direction from SPI toward the IBIS instrument). The
ACS effective area as a function of energy for this direction has
been computed by means of Monte Carlo simulations based on
detailed mass modelling (Weidenspointner et al. 2003, and
references therein) of  SPI  and of the surrounding
material (satellite structure and other instruments). The
effective area increases monotonically with energy; it is 
$\sim$340 cm$^2$ at 100 keV, $\sim$1150 cm$^2$ at 200 keV, and larger
than 3000 cm$^2$ above 1 MeV. Thus the ACS provides the data with
the best statistics in the soft $\gamma$-ray range available for this
giant flare.
For an optically thin thermal bremsstrahlung spectrum with
kT$_{br}$=30 keV   we obtain a conversion factor of 1 ACS count
s$^{-1}$
$\sim$4.3$\times10^{-10}$ erg cm$^{-2}$ s$^{-1}$ (80-2000
keV).

\section{Properties of the giant flare}

The total light curve of the flare, obtained by binning at 2.5 s
the original ACS data, is shown in Fig.~\ref{total}, while
Fig.~\ref{initialfit} displays at full resolution (50 ms) the
initial part of the pulsating tail. We have defined t=0 at
21:30:26.55 UT of 2004 December 27, approximately coinciding with
the rise time of the flare. The giant flare starts with a short
and extremely intense spike\footnote{the Y-axis scale in Fig.~1 is
cut at 1.05$\times$10$^{5}$ counts s$^{-1}$ to better show the
flare tail}. The measured count rate at the peak,
$\sim2\times10^6$ counts s$^{-1}$,  is strongly affected by dead
time and saturation effects which cannot be easily modelled and
prevent reliable flux estimates. A fit with a constant to the
count rate in a 2 ks long time interval before the flare
yields a count rate of
$\sim$8.8$\times$10$^4$ counts s$^{-1}$, which we use in the
following analysis as a constant background level and include in
all the fits keeping its value fixed.

A narrow burst, lasting $\sim$0.2 s, occurs at t=2.8 s (see
Fig.~\ref{initialfit}). Since this delay corresponds to the time
taken by the wavefront to travel from INTEGRAL to the Moon and
back to the satellite, this burst can be explained as
backscattered radiation of the initial spike. A similar detection
was reported with the Helicon-Coronas-F satellite, which could not
observe directly \sgr\ due to Earth occultation (Mazets et al.
2005). Assuming for the backscattered radiation the spectral shape
derived by the above authors (dN/dE$\propto$E$^{-0.7}$ exp(--E/800
keV) ), we obtain a fluence  of $\sim$2$\times10^{-6}$ erg
cm$^{-2}$ (E$>$80 keV). The duration and the fluence of the
backscattered radiation are consistent with the properties of the
initial spike (Mazets et al. 2005; Hurley et al. 2005; Terasawa et
al. 2005).

The inset of Fig.~\ref{total} shows the light curve of the
precursor burst which occurred 143~s before the flare.
Triangulation analysis, using the ACS light curve together with
data from other satellites, demonstrates that its arrival
direction is consistent with the position of \sgr\ (Hurley et al.
2005).
The above authors, using the RHESSI satellite, found that the
precursor spectrum can be crudely approximated by a thermal
bremsstrahlung  with kT$_{br}$=15 keV and reported a fluence of
1.8$\times$10$^{-4}$ erg cm$^{-2}$ for E$>$3 keV. Assuming  the
same spectrum, we obtain a fluence above 80 keV of
4.4$\times$10$^{-6}$ erg cm$^{-2}$, which extrapolated to lower
energy overestimates the RHESSI value by a factor ten. Considering
the uncertainties in the instruments cross calibration and the
extrapolation largely dependent on the  poorly determined
spectrum, we do not consider this discrepancy as particularly
significant. For instance, for kT$_{br}$=30 keV we get F($>$80
keV)=4.0$\times$10$^{-6}$ erg cm$^{-2}$ and F($>$3 keV)=10$^{-4}$
erg cm$^{-2}$. The phase of the event rise time with respect to
the 7.56 s pulsations\footnote{since our data allow to determine
the pulsation period with an accuracy  of only $\sim$0.1 s, we
adopt the more precise value obtained with RXTE (Woods et al.
2005) and define phase $\phi$=0 at t=0 s} is $\phi\sim$0.1, which
does not correspond to any of the peaks seen in the pulsating
tail.

The transition from the initial spike to the pulsating tail is
well fitted, in the time interval 1.2--2.2 s, by a power law
function F$\propto$t$^{-\delta}$ with $\delta$=2.1$\pm$0.1
(Fig.~\ref{initialfit}). The broad bump   in the time range
$\sim$2--5 s, as well as the following ones after t=10 s, are in
phase  with the main peak of the 7.56 s pulsations
($\phi$=0.35--0.6) and can therefore be considered as the first
appearance of the periodicity. The average profile of the
pulsations is shown in Fig.~\ref{fol757-aver} for different time
intervals. An evolution of the profile during the flare, with the
relative intensity of the  peak at $\phi$=0.75 increasing with
time, is evident.

To characterize the long-term decay profile removing the effect of
the pulsations, we  binned the data in time intervals of 7.56 s.
The resulting light curve (inset of Fig.~\ref{initialfit})  is
well fitted in the time interval t$\sim$15--400 s by an
exponential function with decay constant $\tau$=138$\pm$5 s. Note
that the first bin (t=7.56--16.2 s) does not contain the initial
hard spike. Nevertheless it lies significantly above the fitted
function. Assuming a thermal bremsstrahlung spectrum
with kT$_{br}$=30 keV, similar to the giant flares tails of other
SGRs, the fluence in the time interval 1--400 s is
$\sim$2.6$\times$10$^{-4}$ erg cm$^{-2}$ for E$>$80 keV.
Extrapolating the same spectral shape to lower energy, this
corresponds to F($>$3 keV)$\sim$6.4$\times$10$^{-3}$ erg cm$^{-2}$.

\section{Afterglow emission}

Our data show evidence for a second, separate component in
addition to the main flare discussed above. After the end of the
pulsating tail, at t$\sim$400 s, the count rate increases
again, forming a long bump which peaks around t$\sim$600--800 s
and returns to the pre-flare background level at t$\sim$3000-4000
s (see Fig.~\ref{longlongtail}).

The ACS count rate is dominated by the incident flux of charged
particles which varies on different time scales, due to the
satellite motion   and to the temporal and spatial
variations in the flux of cosmic rays and trapped radiation. A
long term ACS light curve covering almost one day and including
the \sgr\ giant flare is shown in the top panel of
Fig.~\ref{psac}. For comparison, the lower panel of
Fig.~\ref{psac} shows the light curve
obtained in the same time interval 
with the Plastic Scintillator Anti-Coincidence (PSAC), a
thin (5 mm) detector located just below the SPI coded mask
aperture, which is sensitive to charged particles and almost
completely transparent to hard X--ray photons. It can be noted
that both the long term trend and most of the fluctuations on
shorter time scales are present in both detectors, as expected for
charged particle induced background. On the contrary, the bump at
t=400--4000 s is prominent only in the ACS, indicating its hard
X--rays nature. Known variable sources, like e.g. Cyg X--1, are
not bright enough in the hard X--rays to produce the count rate
excess seen in the ACS, and no new transients were reported at the
time of this observation. We also checked that the Sun was not
particularly  active in X--rays, by looking at the publicly
available data of the GOES
satellites\footnote{http://www.sec.noaa.gov/}. We therefore
conclude that the re-brightening after t=400 s and lasting about
one hour is associated with \sgr . To our knowledge, this
component has not been reported using data from other satellites.
This might be due to background variability and source occultation
effects in low Earth orbit satellites and to the lack of
sufficient sensitivity at hard X--rays in other detectors.

After accounting for the background, estimated from a linear fit
to the ACS count rate for t$<$0 s and t$>$4000 s, the time profile
decay in the interval t=500-4000 s can be reasonably well
described as a power law F$\propto$t$^{-\delta}$ with
$\delta\sim$0.85. The fluence in the 400-4000 s time interval  is
approximately equal to that contained in the pulsating tail
(t=1--400 s).

\section{Discussion}

The most striking difference between the 2004 December 27 outburst
of \sgr\ and the giant flares previously observed from SGR
0526--66 and SGR 1900$+$14 is the global energetics of the event.
For a distance of 15 kpc (Corbel \& Eikenberry 2004), the hard
X--ray fluence  in the \sgr\ initial spike (Hurley et al. 2005;
Terasawa et al. 2005) implies an isotropic-equivalent energy
release of several 10$^{46}$~d$^{2}_{15}$ erg.
This is at least two orders of magnitude larger than that of the
two other giant flares (1.6$\times10^{44}$ erg for SGR 0526--26;
$>7\times10^{43}$ erg for SGR 1900$+$14). The much higher involved
energy is also reflected in the properties of the transient radio
emission, which reached a luminosity a factor 500 (Cameron et al.
2005, Gaensler et al. 2005) larger than that seen in SGR 1900$+$14
after the 27 August 1998 event (Frail et al. 1999).
On the other hand, the energy output in the \sgr\ pulsating tail
(1.6$\times10^{44}$~d$^{2}_{15}$ erg for E$>$3 keV)  is similar to
that of SGR 1900$+$14 (5$\times10^{43}$ erg) and  of SGR 0526--66
(4$\times10^{44}$ erg). The giant flare of \sgr\ is thus
characterized by a much higher spike-to-tail energy ratio
($\gsim$100)  than the two previous events ($\sim$1). According to
the magnetar model, essentially all the energy release occurs
during the initial $\sim$0.2 s transient phase, when a hot
relativistic fireball is launched. A fraction of this energy is
trapped by closed field lines in the neutron star magnetosphere,
forming an optically thick photon-pair plasma which evaporates
giving rise to the radiation observed in the pulsating tail
(Thompson \& Duncan 1995). The magnetic field strength limits the
amount of energy that can be confined. The fact that this quantity
is similar in the three giant flares, despite the much higher
total energy release of \sgr , is consistent with a magnetic field
of the same order in the three magnetars.

The pulse profile and its time evolution observed with the ACS are
similar to those seen  with RHESSI and Swift (Hurley et al. 2005;
Palmer et al. 2005), which are dominated by photons of lower
energy. Thanks to the large effective area  above 100 keV of the
ACS, we can clearly see variations in the pulse profile at high
energies. The different evolution of the emission components
visible in the pulsating tail (see Fig.~\ref{fol757-aver})
suggests separate emission regions. It is interesting to note that
the pulsations are present since the very early phase of the
flare, as indicated by  the broad pulse visible at t=2--5 s
(Fig.~\ref{initialfit}). This pulse persists at the same
rotational phase in the following cycles, while a secondary peak
at $\phi\sim$0.75 gradually emerges.

While the above results are similar to those obtained with other
satellites, our data provide unique evidence for a second, long
lasting component in the hard X--ray light curve. The emission
after t$\sim$400 s could originate from the neutron star surface
and/or magnetosphere or, alternatively, from the interaction of
the relativistic fireball with the surrounding medium. A search
for pulsations in the ACS data  after t=400 s gave a
negative result, favoring the second interpretation. The afterglow
produced by the \sgr\ giant flare has been observed in the radio
band (Gaensler et al. 2005; Cameron et al. 2005), yielding
estimates of the minimum energy in magnetic field and relativistic
particles, E,  of several 10$^{43}$ erg (see also Nakar, Piran \&
Sari 2005). The hard X--ray fluence in the 400-3000 s time
interval is consistent with this value. With simple gamma-ray
burst afterglow models based on synchrotron emission we can
estimate the bulk Lorentz factor $\Gamma$ from the time t$_{0}$ of
the afterglow onset: $\Gamma\sim$15 (E / 5 10$^{43}$~erg)$^{1/8}$
(n / 0.1 cm$^{-3}$)$^{-1/8}$ (t$_{0}$ / 100 s)$^{-3/8}$, where n
is the ambient density (see, e.g. Zhang \& Meszaros 2003, and
references therein). This is smaller than the typical values for
gamma-ray bursts, but consistent, considering the large involved
uncertainties, with the mildly relativistic outflow inferred from
the modelling of the radio data (Granot et al. 2005).
Alternatively, the observed re-brightening could be due to an
Inverse Compton component, implying a high density environment
(n$\gsim$10 cm$^{-3}$) and a high electron radiation efficiency.

 \acknowledgments

We thank K. Hurley, G. Ghisellini, J. Borkowski and M. Feroci for
useful comments. This work has been partially funded by the
Italian Space Agency. The SPI-ACS is supported by the German
"Ministerium fur Bildung und Forschung" through the DLR grant
50.OG.9503.0.

 \clearpage
 \figcaption{ACS light curve of the whole giant flare binned
at 2.5 s. The peak of the flare, reaching an observed count rate
$\gsim2\times10^{6}$ counts s$^{-1}$, is not shown. The inset
shows the light curve of the precursor burst at full resolution
(50 ms). \label{total}}
 \epsscale{.7}
 \plotone{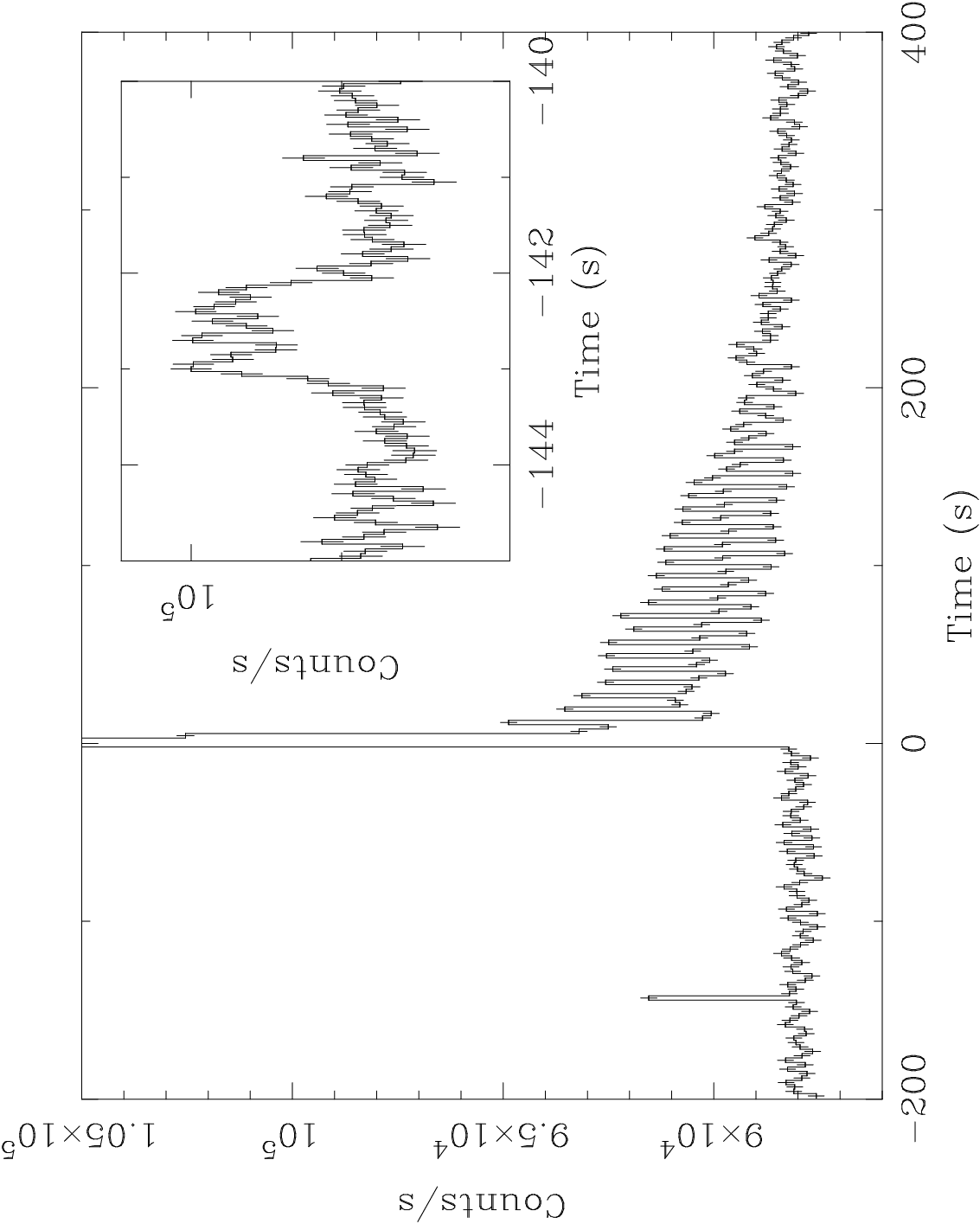}

 \clearpage
 \figcaption{Declining part of the initial spike and initial part of the pulsating tail.
 The line is a fit with a power law plus a constant (fixed at the pre-flare level) to the count rate in the
time interval 1.2--2.2 s. The narrow burst at t=2.8 s is due to
radiation from the initial spike backscattered by the Moon.  The
inset shows the light curve of the tail binned at the spin period
(7.56 s) and fitted with an exponential function.
\label{initialfit}}
 \epsscale{.7}
 \plotone{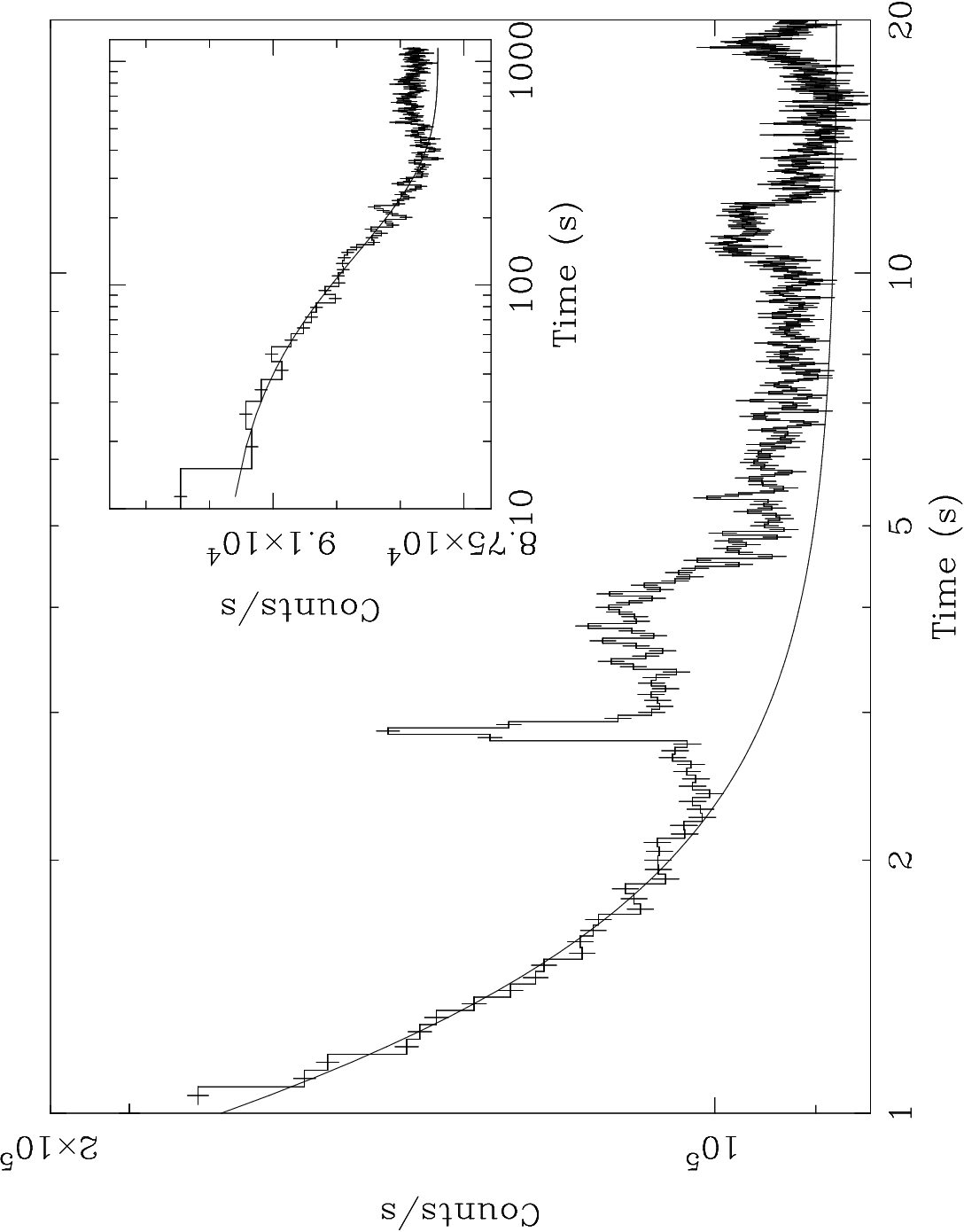}

\clearpage
 \figcaption{Averaged pulse profile obtained by folding the data in three time
intervals at the spin period of 7.56 s.  \label{fol757-aver} }
 \epsscale{.7}
 \plotone{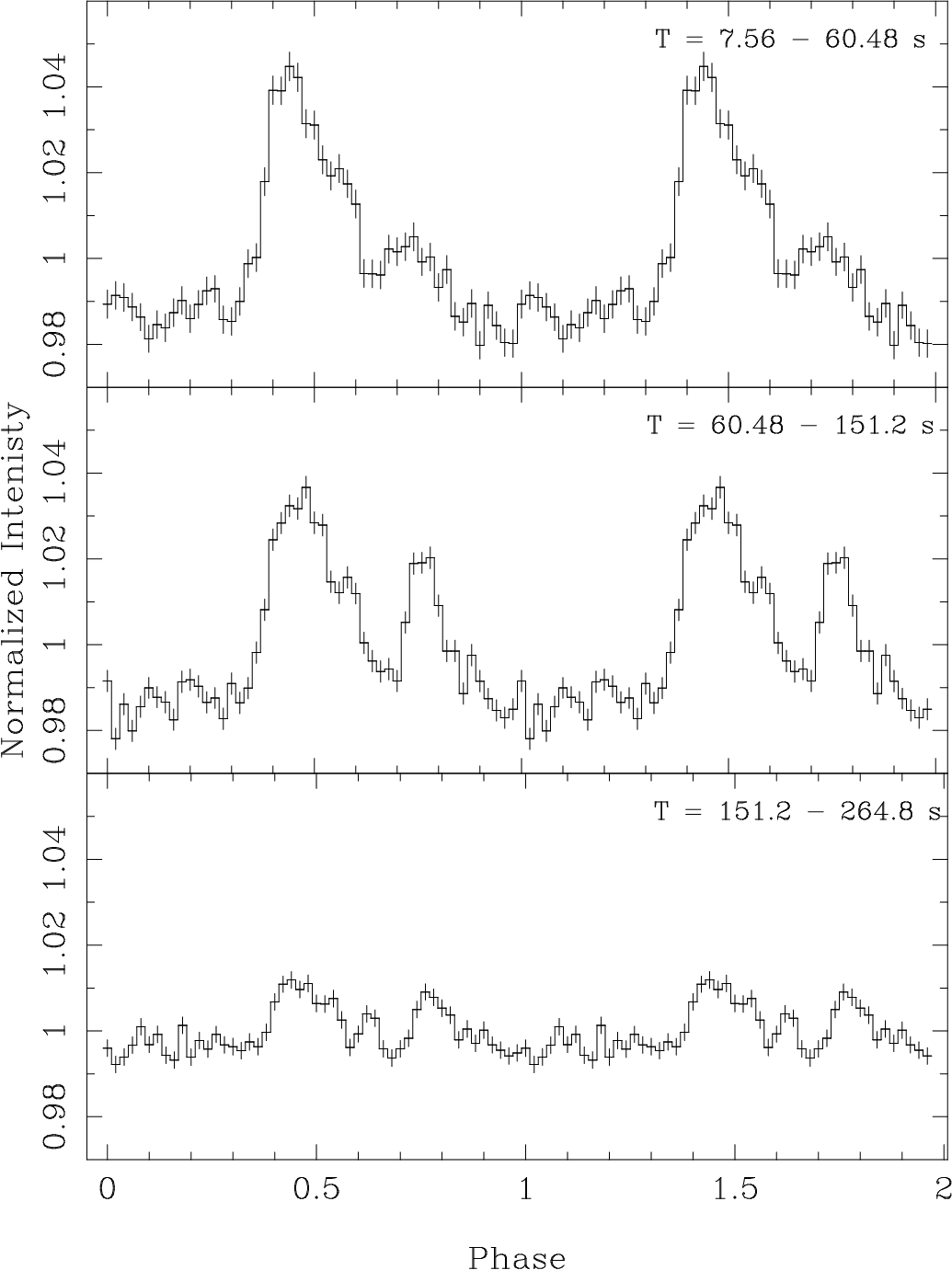}

\clearpage
 \figcaption{SPI ACS light curve of \sgr\ binned at 50 s.  The count rate settles back
to the pre-flare level after t$\sim$3000--4000 s.
 \label{longlongtail}}
 \epsscale{.7}
  \plotone{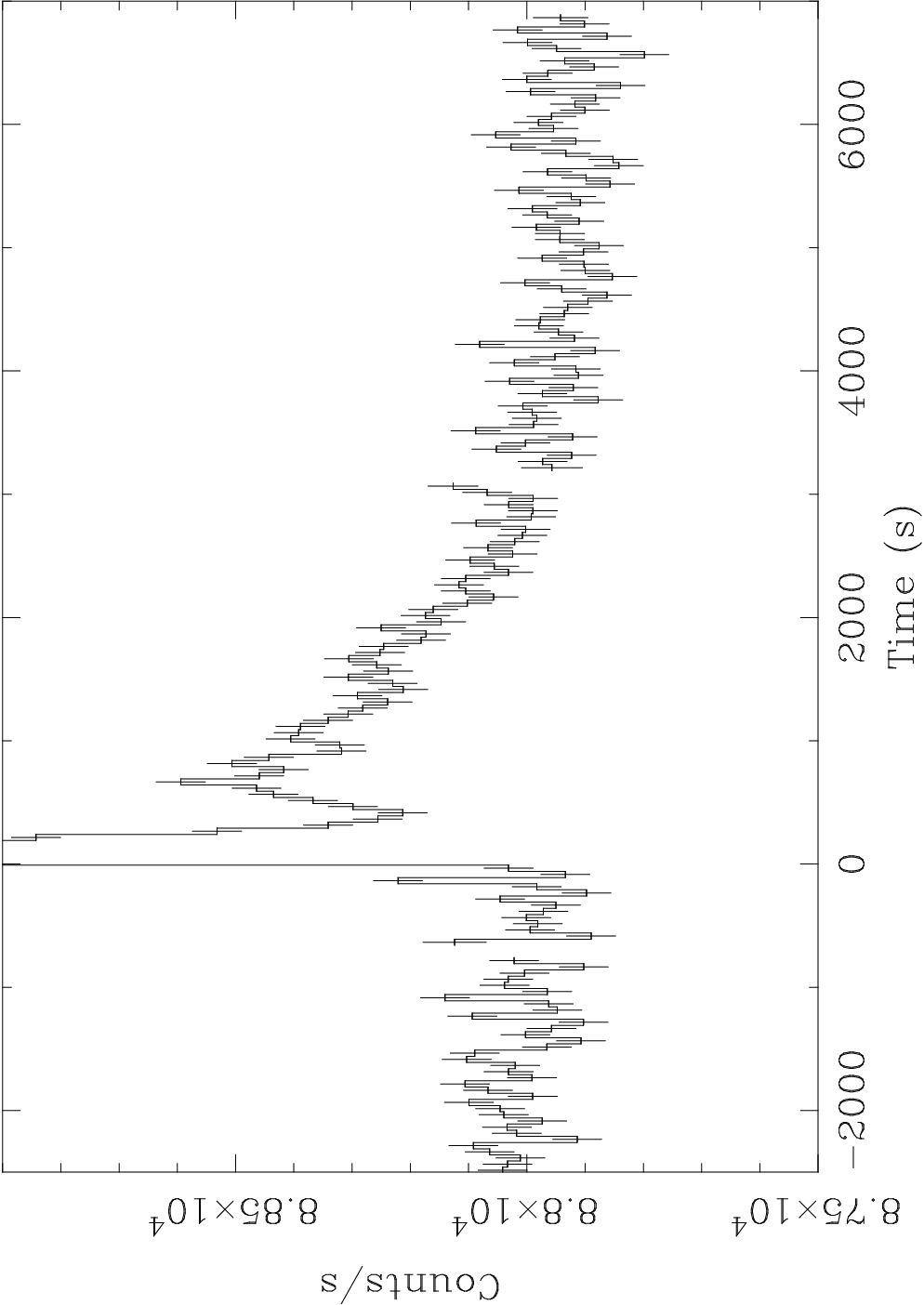}

\clearpage
 \figcaption{ Comparison between the
ACS (top) and PSAC (bottom) light curves. The long lasting
emission after the giant flare is not visible in the PSAC, while
particle induced variations are visible in both detectors (the
narrow pulse at t$\sim-10^4$ s in the ACS is a gamma-ray burst,
unrelated to \sgr ).
  \label{psac}}
 \epsscale{.7}
  \plotone{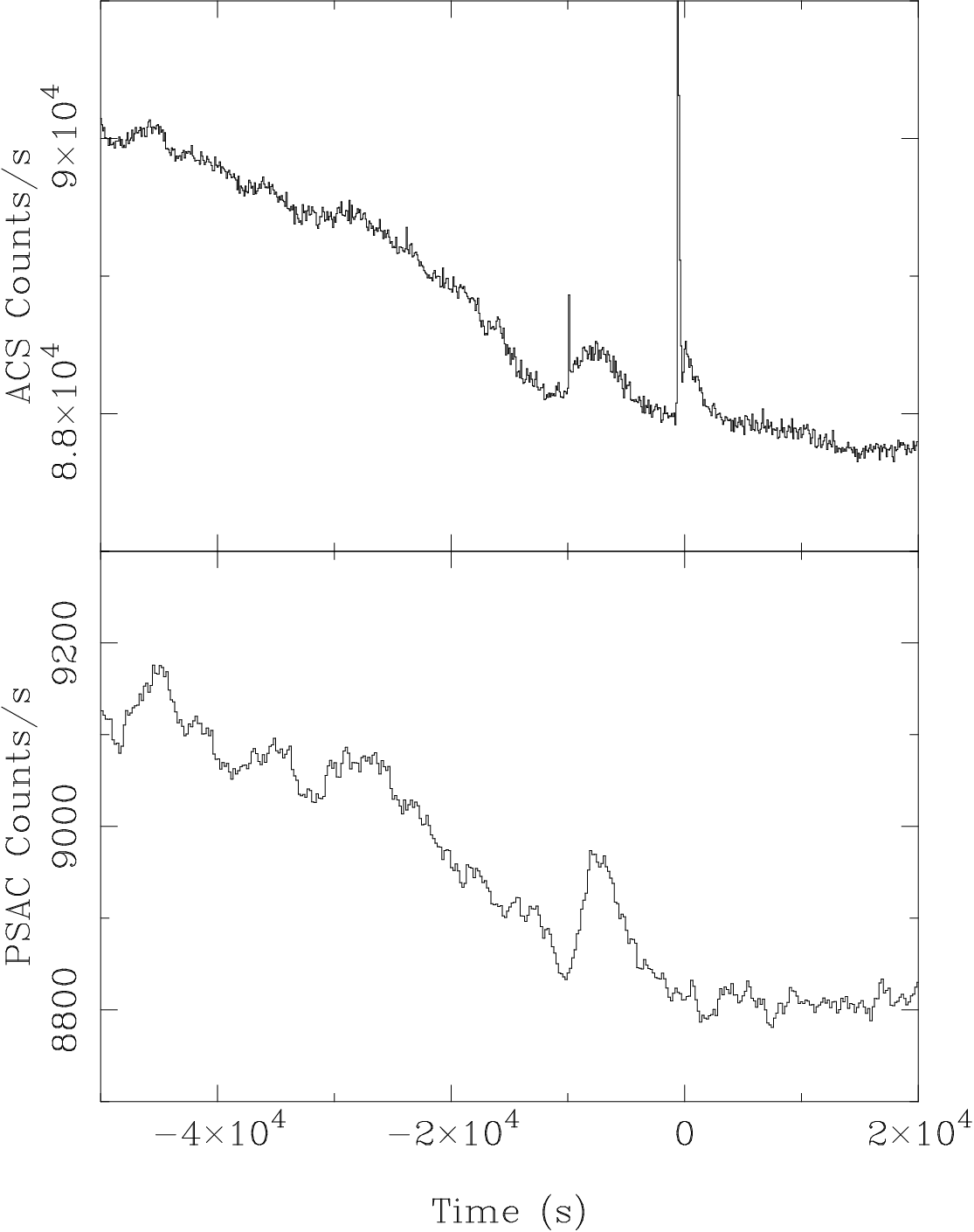}


\begin{references}

 \reference{} Aptekar R.L., Frederiks D.D., Golenetskii S.V. et al. 2001, ApJSS 137, 227
 \reference{} Borkowski J., G\"{o}tz D., Mereghetti S. et al. 2004, GCN Circ. n. 2920
  \reference{} Cameron P.B. et al. 2005, submitted to Nature, astro-ph/0502428
  \reference{} Campbell P. et al. 2005, GCN Circ. n. 2932
  \reference{} Corbel S. \& Eikenberry  S.S. 2004, A\&A 419, 191
 \reference{} Feroci M. et al. 1999, ApJ 515, L9
 \reference{} Frail D.A., Kulkarni, S.R. \& Bloom, J.S., 1999, Nature, 398, 127
 \reference{}Gaensler, B.M., et al., 2005, submitted to Nature, astro-ph/0502393
 \reference{}G{\" o}{\u g}{\" u}{\c s}, E., Kouveliotou, C., Woods, P.M., et al. 2001, ApJ, 558, 228
 \reference {} Golenetskii S.V., Aptekar R.,  Mazets E. et al. 2004, GCN Circular n. 2769
 \reference{}G\"{o}tz, D., Mereghetti S., Mirabel F.I. \& Hurley K.  2004, A\&A 417, L45
 \reference{} Granot J., Ramirez-Ruiz E., Taylor G.B. et al. 2005, Submitted to ApJ, astro-ph/0503251
 \reference{} Hurley K., Cline T., Mazets E. et al.  1999, Nature 397, 41
 \reference{} Hurley K.  et al. 2005, submitted to Nature, astro-ph/0502329
 \reference{} Ibrahim A.I., Swank J.H. \& Parke W. 2003, ApJ 584, L171
 \reference{} Mazets E.P. et al. 1979, Nature 282, 587
 \reference{} Mazets E.P. et al. 2005, astro-ph/0502541
 \reference{} Mereghetti S., G\"{o}tz D., Borkowski J. et al. 2004, GCN Circular n. 2763
 \reference{} Mereghetti S., G\"{o}tz D., Mirabel I.F \& Hurley K. 2005a, A\&A, in press, astro-ph/0411695
 \reference{} Mereghetti S., Tiengo A., Esposito P.  et al. 2005b, submitted to ApJ, astro-ph/0502417
 \reference{} Nakar E., Piran T. \& Sari R. 2005, astro-ph/0502052
 \reference{} Palmer D. et al. 2005, submitted to Nature, astro-ph/0503030
 \reference{} Terasawa, T. et al., 2005, submitted to Nature, astro-ph/0502315
 \reference{}Thompson, C., \& Duncan, R.C. 1995, MNRAS, 275, 255
 \reference{} von Kienlin A., Beckmann V., Rau A., et al. 2003, A\&A, 411, L299
 \reference{} Vedrenne G., Roques J.-P., Sch\"{o}nfelder V., et al. 2003, A\&A 411, L63
 \reference{} Weidenspointner G., Kiener J., Gros M. et al. 2003, A\&A 411, L113
 \reference{} Woods P.M. \& Thompson C. 2004, astro-ph/0406133
 \reference{} Woods P.M., Kouveliotou C., G{\" o}{\u g}{\" u}{\c s} E., et al. 2004, The Astronomer's Telegram, 313
 \reference{} Woods P.M., Finger M., Patel S. et al. 2005, GCN Circular n. 2950
 \reference{} Zhang B. \& M\'{e}sz\'{a}ros P. 2003, astro-ph/0311321


\end{references}
\end{document}